\documentclass[aps,twocolumn,prl,showpacs]{revtex4}
\usepackage{mathrsfs}
\usepackage{graphicx}
\usepackage{amsmath,amssymb,amsthm,bbm,latexsym}
\usepackage{amsthm}
\usepackage{amsbsy}
\usepackage{epsfig}
\usepackage{graphicx}
\usepackage{float}
\usepackage{dcolumn}
\usepackage{amsmath}

\begin{document}

\author{Yong-Cheng Ou,$^1$ Heng Fan,$^1$ and Shao-Ming Fei $^{2,3}$}

\affiliation{$^1$Institute of Physics, Chinese Academy of Sciences,
Beijing 100080, China}

\affiliation{$^{2}$Department of Mathematics, Capital Normal
University, Beijing 100037, China}

\affiliation{$^{3}$Institut f$\ddot{u}$r Angewandte Mathematik,
Universit$\ddot{a}$t Bonn, D-53115, Germany}

\title{Concurrence and a proper monogamy inequality
for arbitrary quantum states }

\begin{abstract}
We obtain an analytical lower bound of entanglement quantified by
concurrence for arbitrary bipartite quantum states. It is shown that
our bound is tight for some mixed states and is complementary to the
previous known lower bounds. On the other hand, it is known that the
entanglement monogamy inequality proposed by Coffman, Kundu, and
Wootters is in general not true for higher dimensional quantum
states. Inducing from the new lower bound of concurrence, we find a
proper form of entanglement monogamy inequality for arbitrary
quantum states.
\end{abstract}

\maketitle

Quantum entanglement is considered to be the most nonclassical
manifestations of quantum mechanics and plays an important role not
only in quantum information sciences but also in condensed-matter
physics \cite{horodecki,vedral}. Due to the decoherence which is in
general unavoidable for quantum system one has to deal with mixed
states in quantum information processing. However, it turns out that
the detection, quantification and distillability of quantum
entanglement for mixed states are much more complicated than
expected though much progress has already been made in the past
years. For example, the operational measure of entanglement for
arbitrary mixed states is still not known, and even the separability
criterion for mixed states can detect all entanglement for only
$2\otimes2$ and $2\otimes3$ systems \cite{p,h}.

The concurrence is one of the well accepted entanglement measure
\cite{wootters}, however, the analytical formulae of concurrence are
only for two-qubit states \cite{wootters} and some high dimensional
bipartite states with certain symmetries, like isotropic ones
\cite{kg}. For general higher dimensional mixed states less has been
known yet \cite{g,mintert} and the optimization method is necessary
and is thus not operational. Instead of analytical exact results,
some operational lower bounds of the concurrence have been derived
recently \cite{cb,chen}, which can detect some bound entangled
states but not all. It is clear that the lower bound of concurrence
can also provide a separability criterion.

On the other hand, the study of distributed entanglement for
multipartite states is important for quantum cryptography
\cite{coffman} and condensed-matter physics \cite{a}. Nevertheless
the monogamy inequality developed by Coffman, Kundu, and Wootters
holds only for qubit systems \cite{coffman,osborne}, it does not
hold in general for higher dimension if we use a straightforward
extension \cite{ou}. It is generally accepted that the entanglement
can not be shared freely, thus a proper definition of the general
monogamy inequality of entanglement is necessary.

In this Letter, the situation above is shown to be dramatically
improved: We derive an analytical lower bound of concurrence for
arbitrary bipartite quantum states by decomposing the joint Hilbert
space into many $2\otimes 2$ dimensional subspaces, which does not
involve any optimization procedure \cite{mintert} and gives an
effective evaluation of entanglement together with an operational
sufficient condition for the distillability of any bipartite quantum
states, which improves the result \cite{ho1}. Inducing from our
bound, for the first time, we generalize the monogamy inequality
developed by Coffman, Kundu, and Wootters to any pure multipartite
quantum states. This proper inequality not only is the fundamental
constraint for entanglement sharing but also can define a measure of
multipartite entanglement.

For a pure bipartite state $\rho_{AB}=|\psi\rangle\langle\psi|$ in a
finite $d_1\otimes d_2$ dimensional Hilbert space $\mathcal{H}
_A\otimes \mathcal{H}_B$, the concurrence is defined as $\mathcal
{C}(|\psi\rangle)=\sqrt{2\left(1-\texttt{Tr}\rho_A^2\right)}$ with
$\rho_A=\texttt{Tr}_B\rho_{AB}$ the reduced density matrix. A pure
state can be generally expressed as
$|\psi\rangle=\sum_{i=1}^{d_1}\sum_{j=1}^{d_2}\phi_{ij}|ij\rangle$,
$\phi_{ij}\in$ $\mathbf{C}$, in computational basis $|i\rangle$ and
$|j\rangle$ of $\mathcal{H}_A$ and $\mathcal{H}_B$ respectively,
$i=1,...,d_1$ and $j=1,...,d_2$. After some algebraic calculations
\cite{fanfei}, one derives the squared concurrence:
\begin{equation}\label{aa}
\mathcal
{C}^2(|\psi\rangle)=\sum_{m=1}^{D_1}\sum_{n=1}^{D_2}|\mathcal
{C}_{mn}|^2=4\sum_{i<j}^{d_1}\sum_{k<l}^{d_2}|\phi_{ik}\phi_{jl}-\phi_{il}\phi_{jk}|^2,
\end{equation}
where $D_1=d_1(d_1-1)/2$, $D_2=d_2(d_2-1)/2$, $\mathcal
{C}_{mn}=\langle\psi|\widetilde{\psi}_{mn}\rangle$,
$|\widetilde{\psi}_{mn}\rangle=(L_{m}\otimes L_{n})|\psi^*\rangle$,
and $L_m, m=1,..., d_1(d_1-1)/2$, $L_n, n=1, ..., d_2(d_2-1)/2$ are
the generators of group $SO(d_1)$ and $SO(d_2)$ respectively.

From Eq.(\ref{aa}) it is evident that the $d_1\otimes d_2$
dimensional Hilbert space is decomposed into
$d_1(d_1-1)d_2(d_2-1)/4$ $2 \otimes 2$ dimensional subspaces, such
that the squared concurrence is just the sum of all squared
``two-qubit'' concurrences. A pure state is separable iff all these
``two qubits" are separable. For a mixed state $\rho$
\begin{equation}\label{rho}
\rho=\sum_{i}p_i|\psi_i\rangle\langle\psi_i|,~~
p_i\geq 0,~~\sum_i p_i=1,
\end{equation}
the concurrence is defined by the convex-roof:
\begin{eqnarray}\label{b}
\mathcal {C}(\rho)\equiv \min\sum_{i}p_i\mathcal {C}(|\psi_i\rangle)
\end{eqnarray}
of all possible decompositions into the pure states
$|\psi_i\rangle$. Although the concurrence (\ref{b}) is cumbersome
to solve due to a high dimensional optimization, one may provide an
analytical lower bound on it as shown in the following.

\textbf{Theorem 1:} For an arbitrary $d_1\otimes d_2$ state (\ref{rho}),
the concurrence $\mathcal {C}(\rho)$ satisfies
\begin{equation}\label{c}
 \tau(\rho)\equiv \sum_{m=1}^{d_1(d_1-1)/2}\sum_{n=1}^{d_2(d_2-1)/2} \mathcal
  {C}_{mn}^2\leq \mathcal {C}^2(\rho),
\end{equation}
where $\tau$ is a lower bound of squared concurrence and
\begin{equation}\label{d}
\mathcal {C}_{mn}=\max{\{0,
\lambda_{mn}^{(1)}-\lambda_{mn}^{(2)}-\lambda_{mn}^{(3)}-\lambda_{mn}^{(4)}\}},
\end{equation}
with $\lambda^{(1)}_{mn},...,\lambda^{(4)}_{mn}$ being the square
roots of the four nonzero eigenvalues, in decreasing order, of the
non-Hermitian matrix $\rho\widetilde{\rho}_{mn}$ where
$\widetilde{\rho}_{mn}=(L_{m}\otimes L_{n})\rho^*(L_{m}\otimes
L_{n})$.

\textbf{Proof:} Set $|\xi_i\rangle=\sqrt{p_i}|\psi_i\rangle$. The
concurrence (\ref{b}) takes the form:
\begin{equation}\label{c5}
\mathcal {C}(\rho)=\min\sum_i\left(
\sum_{m=1}^{d_1(d_1-1)/2}\sum_{n=1}^{d_2(d_2-1)/2}|\langle\xi_i|
L_m\otimes L_n|\xi_i^*\rangle|^2\right)^{\frac{1}{2}}.
\end{equation}
For simplicity, we denote the term after the min as $\mathcal {D}$,
i.e., $\mathcal {C}(\rho)=\min\mathcal {D}$. Recall that for any
function $F=\sum_i\left(\sum_j x_{ij}^2\right)^{1/2}$ subjected to
the constraints $z_j=\sum_i x_{ij}$ with $x_{ij}$ being real and
nonnegative, the inequality $\sum_j z_j^2\leq F^2$ holds, from which
it follows that $\mathcal {D}$ satisfies
\begin{equation}\label{c8}
\sum_{m=1}^{d_1(d_1-1)/2}\sum_{n=1}^{d_2(d_2-1)/2}\left(\sum_i|\langle\xi_i|
L_m\otimes L_n|\xi_i^*\rangle|\right)^2\leq \mathcal {D}^2.
\end{equation}
In order to seek a lower bound of the minimum of $\mathcal {D}$ over
all pure-state decompositions, we only need to consider
\begin{equation}\label{c9}
\mathcal {C}^{'}_{mn}=\min\sum_i|\langle\xi_i| L_m\otimes
L_n|\xi_i^*\rangle|,
\end{equation}
for all $m$ and $n$, by using procedure of extremizations adopted in
\cite{wootters, mintert}. Let $\lambda_i$ and $|\chi_i\rangle$ be
eigenvalues and eigenvectors of $\rho$ respectively. Any
decomposition of $\rho$ can be obtained from a unitary $d\times d$
matrix $V_{ij}$,
$|\xi_{j}\rangle=\sum_{i=1}^{d}V^*_{ij}(\sqrt{\lambda_{i}}|\chi_{i}\rangle)$.
Therefore one has $\langle\xi_i| L_m\otimes
L_n|\xi_j^*\rangle=(VYV^T)_{ij}$, where the matrix $Y$ is defined by
$Y_{ij}=\langle \chi_i| L_m\otimes L_n|\chi_j^*\rangle$.
Eq.(\ref{c9}) turns out to be $\mathcal
{C}^{'}_{mn}=\min\sum_i|[VYV^T]_{ii}|=\lambda_{mn}^{(1)}-\sum_{i>1}\lambda_{mn}^{(i)}$\cite{wootters,
um}, where $\lambda_{mn}^{(j)}$ are the square roots of the
eigenvalues of the positive Hermitian matrix $YY^\dagger$, or
equivalently the non-Hermitian matrix $\rho\widetilde{\rho}_{mn}$ in
decreasing order. As the matrix $L_m\otimes L_n $ has $d_1d_2-4$
rows and $d_1d_2-4$ columns that are identically zero, the matrix
$\rho\widetilde{\rho}_{mn}$ has a rank no greater than 4, i.e.,
$\lambda_{mn}^{(j)}=0$ for $j\geq 5$. From the above analysis we
have Eqs.(\ref{c})-(\ref{d}). $\hfill$$\Box$

Remark: Our bound $\tau$ (\ref{c}) in fact characterizes some
``two-qubit'' entanglement in a high dimensional bipartite state.
One can directly verify that there are at most $4\times 4=16$
nonzero elements in each matrix $\widetilde{\rho}_{mn}$ so as to
lead to a $4\times
4$ matrix $\varrho(\sigma_y\otimes \sigma_y)\varrho^*(\sigma_y%
\otimes \sigma_y)$,  where $\sigma_y$ is the Pauli matrix and the
matrix $\varrho$ is a submatrix of the original $\rho$:
\begin{equation}\label{c11}
\varrho=\left(\begin{array}{cccc}
               \rho_{ik,ik} &  \rho_{ik,il} &  \rho_{ik,jk} &  \rho_{ik,jl}
               \\
               \rho_{il,ik} &  \rho_{il,il} &  \rho_{il,jk} &  \rho_{il,jl}
               \\
               \rho_{jk,ik} &  \rho_{jk,il} &  \rho_{jk,jk} &  \rho_{jk,jl}
               \\
               \rho_{jl,ik} &  \rho_{jl,il} &  \rho_{jl,jk} &  \rho_{jl,jl} \\
             \end{array}\right),
\end{equation}
$i\neq j$ and $k \neq l$, with subindices $i$ and $j$ associated
with the space $\mathcal{H}_A$, and $k$ and $l$ with the space
$\mathcal{H}_B$. The ``two-qubit'' submatrix $\varrho$ is not
normalized but positive semidefinite, such that $\mathcal {C}_{mn}$
is just the concurrence of the state (\ref{c11}).

Our bound $\tau$ also provides a much clearer structure of
entanglement, which not only yields an effective separability
criterion and an easy evaluation of entanglement, but also helps one
to classify mixed-state entanglement. Based on the positive partial
transpose (PPT) criterion, a necessary and sufficient condition for
the distillability was proposed in \cite{ho1}, which is not
operational in general. In the following, we derive an alternative
distillability criterion based on our bound $\tau$ to improve the
operationality to some degree.

\textbf{Theorem 2:} For any bipartite quantum state $\rho$, if
$\tau(\rho^{\otimes N})>0$ for a certain positive integer $N$,
$\rho$ is distillable.

\textbf{Proof:} It was shown in \cite{ho1} that a density matrix
$\rho$ is distillable iff there are some projectors $A$, $B$ that
map high dimensional spaces to two-dimensional ones and certain
number $N$ such that the state $A\otimes B\rho^{\otimes N}A\otimes
B$ is entangled. Thus if $\tau(\rho^{\otimes N})>0$, there exists
one submatrix of matrix $\rho^{\otimes N}$ similar as (\ref{c11})
which has non-zero $\tau $ and is entangled in $2\otimes 2$ space.
So we know that $\rho $ is distillable. We remark that this
submatrix which has a positive $\tau $ is the entangled state
$A\otimes B\rho^{\otimes N}A\otimes B$ up to normalization.
$\hfill$$\Box$

\textbf{Corollary 1:} The lower bound $\tau(\rho)>0$ is a sufficient
condition for the distillability of any bipartite state $\rho$.

\textbf{Corollary 2:} The lower bound $\tau(\rho)=0$ is a necessary
condition for the separability of any bipartite state $\rho$.

Remark: Corollary 1 directly follows from Theorem 2 and this case is
referred as 1-distillable \cite{wd}. The opposite direction of
Theorem 2, whether $\tau(\rho^{\otimes N})>0$ is a necessary
condition of distillability is still a challenging question. The
answer to this question can not only provide an operational
criterion for mixed states distillability but also shed light on
question of whether the non-PPT (NPPT) nondistillable states exist
which is studied numerically in \cite{dp,wd}.

Our bound $\tau$, PPT criterion, separability and distillability for
any bipartite quantum state $\rho$ have the following relations. If
$\tau(\rho)>0$, $\rho$ is entangled. If $\rho$ is separable, it is
PPT. If $\tau(\rho)>0$, $\rho$ is distillable. If $\rho$ is
distillable, it is NPPT. From the last two propositions it follows
that if $\rho$ is PPT, $\tau(\rho)=0$, i.e., if $\tau(\rho)>0$,
$\rho$ is NPPT. We give some examples below.

\emph{Example 1:} Horodecki's $3\otimes3$ system \cite{h5}:
\begin{equation}\label{c14}
\begin{array}{cc}
  \sigma_\alpha=\frac{2}{7}|\Psi^+\rangle\langle\Psi^+|+\frac{\alpha}{7}\sigma_++\frac{5-\alpha}{7}\alpha_-,
 &
\end{array}
\end{equation}
where
$\sigma_+=\frac{1}{3}(|01\rangle\langle01|+|12\rangle\langle12|+|20\rangle\langle20|)$,
$\sigma_-=\frac{1}{3}(|10\rangle\langle10|+|21\rangle\langle21|+|02\rangle\langle02|)$,
and $|\Psi^+\rangle$ is a maximally entangled state. The state
$\sigma_\alpha$ is separable for $2\leq\alpha\leq3$; bound entangled
for $3<\alpha\leq4$; free entangled for $4<\alpha\leq5$ \cite{h5}.
From lower bound $\tau$ in (\ref{c}) we have $\tau(\sigma_\alpha)=0$
for $2\leq\alpha\leq4$ and
$\tau(\sigma_\alpha)=4\left(2-\sqrt{\alpha(5-\alpha)}\right)^2/147$
for $4<\alpha\leq5$. According to Corollary 1 since
$\tau(\sigma_\alpha)>0$ for $4<\alpha\leq5$, the state is
distillable, agreeing with the conclusion in \cite{h5}. Note that
our lower bound is weaker than the one by realignment\cite{chen} for
$4\leq \alpha \lesssim4.79$ but stronger for $4.79\lesssim \alpha
 \leq 5$.

\emph{Example 2:} Isotropic states in $d\otimes d$
dimensions \cite{h6, kg}:
\begin{equation}\label{c16}
\rho_{F}=\frac{1-F}{d^2-1}\left(\emph{I}-|\Phi^+\rangle\langle\Phi^+|\right)+F|\Phi^+\rangle\langle\Phi^+|,
\end{equation}
where $|\Phi^+\rangle$ is a maximally entangled state. These states
are separable for $F\leq1/d$ \cite{h6}. Our bound $\tau$ gives
$\tau(\rho_F)=0$ for $F\leq1/d$, and $\tau(\rho_F)=2(dF-1)^2/d(d-1)$
for $F>1/d$, which is just the exact squared concurrence \cite{rt}.
Thus these states saturate the inequality (\ref{c}), which implies
that the entanglement of these states is composed of only the
entanglement of the``two qubits'' in each state. According to
Corollary 1 since $\tau(\rho_F)>0$ for $F>1/d$, all these states are
distillable, agreeing with the analysis in \cite{ho1}.

Now we show a NPPT quantum state with $\tau=0$.

\emph{Example 3:} Werner states in $3\otimes3$ dimensions \cite{wern}:

\begin{equation}\label{c17}
\rho_W(\lambda)=\frac{1}{8\lambda-1}\left(\lambda
\emph{I}-\frac{\lambda+1}{3}H\right),
\end{equation}
where $H|i,j\rangle=|j,i\rangle$ for all $i, j=1, 2, 3$. For any
finite $\lambda > 0$ the state $\rho_W(\lambda)$  is a NPPT state
\cite{shor}. It is conjectured that for $\lambda\geq 2$ the state
$\rho_W(\lambda)$ is undistillable \cite{dp, wd}. Our lower bound
$\tau$ of the state (\ref{c17})
$\tau(\rho_W(\lambda))=(4-2\lambda)^2/3(8\lambda-1)^2>0$ for
$0<\lambda <2$. Hence these states in this parameter region are
distillable according to Corollary 1, agreeing with the analysis in
\cite{wd}. While the lower bound is $\tau(\rho_W(\lambda))=0$ for
$\lambda \geq2$, these states are just the NPPT states with $\tau=0$
and they are 1-copy undistillable. However, the nondistillability of
$N$-copy dose not imply the undistillability of $N+1$-copy
\cite{ww}. And the technique in Theorem 2 may help to confirm the
long-standing conjecture by computing $\tau(\rho^{\otimes N})$.

From these examples one can explicitly see that our bound $\tau$
provides an easy evaluation of concurrence for most of the free
entangled states. On the other hand, as we known, the entanglement
is monogamous \cite{coffman}, however, surprisingly, a direct
extension of the monogamy inequality from qubit case to the general
case does not work \cite{ou,horodecki}. In this Letter,
interestingly, one fundamental property of our bound $\tau$ is shown
that it is monogamous also. It is thus a proper definition of the
monogamy inequality for general cases.

\textbf{Theorem 3:} For any pure tripartite state
$|\psi\rangle_{ABC}$ in arbitrary $d_1\otimes d _2\otimes d_3$
dimensional spaces, the lower bound $\tau$ of concurrence satisfies
\begin{equation}\label{c18}
\tau(\rho_{AB})+\tau(\rho_{AC})\leq\tau(\rho_{A:BC}),
\end{equation}
where $\rho_{AB}=\texttt{Tr}_{C}(|\psi\rangle_{ABC}\langle\psi|)$,
$\rho_{AC}=\texttt{Tr}_{B}(|\psi\rangle_{ABC}\langle\psi|)$, and
$\rho_{A:BC}=\texttt{Tr}_{BC}(|\psi\rangle_{ABC}\langle\psi|)$.

\textbf{Proof:} Since $\mathcal {C}_{mn}^2\leq
\left(\lambda^{(1)}_{mn}\right)^2\leq\sum^{4}_{i=1}\left(\lambda^{(i)}_{mn}\right)^2=
\texttt{Tr}(\rho\widetilde{\rho}_{mn})$ for each $m$ and $n$ in
Eqs.(\ref{c}) and (\ref{d}), one can derive the inequality:
\begin{eqnarray}
  \tau(\rho_{AB})+\tau(\rho_{AC})&\leq & \sum_l^{D_1}
  \sum_k^{D_2}\texttt{Tr}\left[\rho_{AB}(\widetilde{\rho}_{AB})_{lk}\right] \notag\\
 &+&
 \sum_p^{D_1}\sum_q^{D_3}\texttt{Tr}\left[\rho_{AC}(\widetilde{\rho}_{AC})_{pq}\right],
\end{eqnarray}
where $D_3=d_3(d_3-1)/2$. By using a similar analysis in
\cite{coffman} one has
$\sum_{lk}\texttt{Tr}\left[\rho_{AB}(\widetilde{\rho}_{AB})_{lk}\right]=
1-\texttt{Tr}\rho_A^2- \texttt{Tr}\rho_B^2+\texttt{Tr}\rho_C^2$ and
$\sum_{pq}\texttt{Tr}\left[\rho_{AC}(\widetilde{\rho}_{AC})_{pq}\right]=
1-\texttt{Tr}\rho_A^2+\texttt{Tr}\rho_B^2-\texttt{Tr}\rho_C^2$,
where $l,p=1,...,D_1$, $k=1,...,D_2$, $q=1,...,D_3$. The sum of
these two inequalities results in that the right-hand side of
(\ref{c19}) equals to $ 2(1-\texttt{Tr}\rho^2_A)=\mathcal
{C}^2(\rho_{A:BC})$. Taking into account that
$\tau(\rho_{A:BC})=\mathcal {C}^2(\rho_{A:BC})$ for a pure state,
one obtains the inequality (\ref{c18}). $\hfill$$\Box$

\textbf{Corollary 3:} Subsequently, we have the general monogamy
inequality
\begin{equation}\label{c19}
\tau(\rho_{AB_1})+\tau(\rho_{AB_2})+...+\tau(\rho_{AB_n})\leq\tau(\rho_{A:B_1...B_n}),
\end{equation}
for any pure multipartite quantum state $\rho_{AB_1B_2...B_n}$ and
$A, B_1, ..., B_n$ may contain any number of particles,
respectively.

To see the tightness of the inequality (\ref{c18}), we consider the
following examples.

\emph{Example 4:} Aharonov state of three qutrits:
\begin{equation}
|\psi\rangle_{ABC}=\frac{1}{\sqrt{6}}(|012\rangle+|120\rangle+|201\rangle-|021\rangle-|102\rangle-|210\rangle).
\nonumber
\end{equation}
For this state, it was shown that the original
Coffman-Kundu-Wootters inequality is violated since $\mathcal
{C}^2(\rho_{AB})+\mathcal {C}^2(\rho_{AC})=2>\mathcal
{C}^2(\rho_{A:BC})=\frac{4}{3}$ \cite{ou}. According to Theorem 3 we have
$\tau(\rho_{AB})+\tau(\rho_{AC})=\frac{2}{3}<\tau(\rho_{A:BC})=\frac{4}{3}$,
thus the inequality (\ref{c18}) is satisfied.

\emph{Example 5:} Generalized five-qubit $W$ state:
\begin{eqnarray}
|\phi\rangle_{ABC}&=&\alpha|10000\rangle+\beta|01000\rangle+\gamma|00100\rangle
\nonumber \\
&&+\delta|00010\rangle+\eta|00001\rangle, \label{c21}
\end{eqnarray}
where the subsystem $B$ (resp. $C$) contains the second and third
(resp. the last two) qubits. The $ABC$ system is $2\otimes 4\otimes
4$ dimensional. For this state, one finds that
$\tau({\rho_{AB}})=4|\alpha|^2(|\beta|^2+|\gamma|^2)$,
$\tau({\rho_{AC}})=4|\alpha|^2(|\delta|^2+|\eta|^2)$, and
$\tau({\rho_{A:BC}})=4|\alpha|^2(|\beta|^2+|\gamma|^2+|\delta|^2+|\eta|^2)$.
Thus $\tau(\rho_{AB})+\tau(\rho_{AC})=\tau(\rho_{A:BC})$, i.e., the
state (\ref{c21}) saturates the inequality (\ref{c18}).

Due to the monogamy inequality, the difference between the two sides
of (\ref{c18}) can be interpreted as a residual entanglement
$\tau_{ABC}$ as for the qubit case \cite{coffman}:
\begin{equation}\label{c20}
\tau_{ABC}=\tau(\rho_{A:BC})- \tau(\rho_{AB})-\tau(\rho_{AC}),
\end{equation}
which, as usual, can be served as a measure of multipartite
entanglement and is fully analytical. Moreover a pure tripartite
state $|\psi\rangle_{ABC}$ can be expressed in the standard basis
$\{|ijk\rangle\}$, where $i=1,...,d_1, j=1, ...,d_2,$ and $k=1,...,
d_3$:
$|\psi\rangle_{ABC}=\sum_{i=1}^{d_1}\sum_{j=1}^{d_2}\sum_{k=1}^{d_3}\phi_{ijk}|ijk\rangle$.
Now the whole joint Hilbert space $\mathcal{H}_A\otimes
\mathcal{H}_B\otimes \mathcal{H}_C$ can be decomposed into
$d_1(d_1-1)d_2(d_2-1)d_3(d_3-1)/8$ $2\otimes 2\otimes 2$ dimensional
subspaces. While each $2\otimes 2\otimes 2$ dimensional subspace has
a same form of the residual entanglement as the one \cite{coffman}
in terms of the coefficients $\phi_{ijk}$. Therefore the residual
entanglement (\ref{c20}) for any pure tripartite state takes the
expression:
\begin{equation}\label{c26}
\tau_{ABC}=4\sum_{i,i'=1}^{d_1}\sum_{j,j'=1}^{d_2}\sum_{k,k'=1}^{d_3}|d_{ijk}^{(1)}-2d_{ijk}^{(2)}+4d_{ijk}^{(3)}|,
\end{equation}
where
\begin{equation}\label{c23}
\begin{array}{c}
  d_{ijk}^{(1)}=\phi_{ijk}^2\phi_{i'j'k'}^2+\phi_{ijk'}^2\phi_{i'j'k}^2+\phi_{ij'k}^2\phi_{i'jk'}^2+\phi_{i'jk}^2\phi_{ij'k'}^2, \\
  d_{ijk}^{(2)}= \phi_{ijk}\phi_{i'j'k'}\phi_{ij'k'}\phi_{i'jk}+\phi_{ijk}\phi_{i'j'k'}\phi_{i'jk'}\phi_{ij'k}\\
  +\phi_{ijk}\phi_{i'j'k'}\phi_{i'j'k}\phi_{ijk'}+\phi_{ij'k'}\phi_{i'jk}\phi_{i'jk'}\phi_{ij'k} \\
  +\phi_{ij'k'}\phi_{i'jk}\phi_{i'j'k}\phi_{ijk'}+\phi_{i'jk'}\phi_{ij'k}\phi_{i'j'k}\phi_{ijk'},\\
  d_{ijk}^{(3)}=\phi_{ijk}\phi_{i'j'k}\phi_{i'jk'}\phi_{ij'k'}+\phi_{i'j'k'}\phi_{ijk'}\phi_{ij'k}\phi_{i'jk},
\end{array}
\end{equation}
with the constraints of the subindices  $i<i', j<j'$ and $k<k'$.
Since each $d_{ijk}^{(l)}$ is symmetrical with respect to $i,j$ and
$k$, it is invariant under permutations of the subsystems $A$, $B$
and $C$. Thus the residual entanglement $\tau_{ABC}$ (\ref{c26}) is
invariant under such permutations.

In summary, we have shown a novel lower bound of concurrence for any
bipartite quantum states, which can be analytically obtained by
calculating all ``two-qubit'' concurrences and is complementary to
known results. Our bound becomes exact for some mixed states. It is
an operational sufficient criterion for the distillability. With the
form of the lower bound,  the monogamy inequality developed by
Coffman, Kundu, and Wootters is generalized to any pure multipartite
quantum states. Consequently one can define a measure of
multipartite entanglement, which can find wide potential
applications in studying quantum phase transition \cite{osterloh,wu}
and in seeking the ground-state energy of condensed-matter systems
\cite{f}. The method developed might also help to calculate the
entanglement of formation and the distillation rate of entanglement.

\bigskip

Y.C.O. was supported by the Postdoctoral Science Foundation of China
and the NSFC (60578055). H.F. was supported by the `Bairen' program
NSFC and the `973' program (2006CB921107). S.M.F. was supported by
the NKBRPC (2004CB318000) and the NSFC (10675086).

\end{document}